\documentclass[a4paper, twosides]{article}
\usepackage{CJK,multicol,multirow,graphics,fancyhdr,epstopdf}
\usepackage{amsmath,amsfonts,amssymb,bm,upgreek,mathrsfs,ccmap,mathcomp}
\usepackage[pagewise,switch,columnwise]{lineno}
\usepackage[compress,nospace]{cite}
\usepackage[dvipsnames]{xcolor}
\usepackage{CPL-2022}
\usepackage{float}
\usepackage{graphicx}
\usepackage{threeparttable}
\usepackage{booktabs}
\usepackage{CJK}

\newcommand{\cplyear}{2022} \newcommand{\cplvol}{39}
\newcommand{\cplno}{x} \newcommand{\cplpagenumber}{xxxxxx}
 \newcommand{\cplpage}{\cplpagenumber-\thepage}

\begin{document}

\begin{CJK}{GBK}{song}\vspace* {-4mm} \begin{center}
\large\bf{\boldmath{Pathfinding pulsar observations with the CVN incorporating the FAST }}
\footnotetext{\hspace*{-5.4mm}$^{*}$Corresponding authors. Email: yanzhen@shao.ac.cn

\noindent\copyright\,{\cplyear}
\href{http://www.cps-net.org.cn}{Chinese Physical Society} and
\href{http://www.iop.org}{IOP Publishing Ltd}}
\\[5mm]
\normalsize \rm{}Zhen Yan$^{1,5}$, Zhiqiang Shen$^{1,5}$, Peng Jiang$^{2,5}$, Bo Zhang$^{1,5}$,  Haiyan Zhang$^{2,5}$, Lang Cui$^{3,5}$, Jintao Luo$^{4,5}$, Rurong Chen$^{2}$, Wu Jiang$^{1,5}$, Hua Zhang$^{3}$, De Wu$^{4}$, Rongbing Zhao$^{1,5}$, Jianping Yuan$^{3,5}$, Yue Hu$^{4}$, Yajun Wu$^{1,5}$, Bo Xia$^{1}$, Guanghui Li$^{3}$, Yongnan Rao$^{4,5}$,  Chenyu Chen$^{3}$, Xiaowei Wang$^{1,5}$, Hao Ding$^{1}$, Yongpeng Liu$^{4}$, Fuchen Zhang$^{4}$, Yongbin Jiang$^{1}$
\\[3mm]\small\sl $^{1}$Shanghai Astronomical Observatory, Chinese Academy of Sciences, Shanghai 200030, China

$^{2}$National Astronomical Observatories, Chinese Academy of Sciences, Beijing 100101, China

$^{3}$Xinjiang Astronomical Observatory, Chinese Academy of Sciences, Urumqi 830011, China

$^{4}$National Time Service Center, Chinese Academy of Sciences, Xi'an 710600, China

$^{5}$School of Astronomy and Space Science, University of Chinese Academy of Sciences, Beijing 100049, China
\\[4mm]\normalsize\rm{}(Received xxx; accepted manuscript online xxx)
\end{center}
\end{CJK}
\vskip 1.5mm

\small{\narrower The importance of Very Long Baseline Interferometry (VLBI) for pulsar research is becoming increasingly prominent and receiving more and more attention. In this paper, we present pathfinding pulsar observation results with the Chinese VLBI Network (CVN) incorporating the Five-hundred-meter Aperture Spherical radio Telescope (FAST). On MJD~60045 (April 11th, 2023), PSRs~B0919+06 and B1133+16 were observed with the phase-referencing mode in the L-band using four radio telescopes (FAST, TianMa, Haoping and Nanshan) and correlated with the pulsar binning mode of the distributed FX-style software correlator in Shanghai. After further data processing with the NRAO Astronomical Image Processing System (AIPS), we detected these two pulsars and fitted their current positions with accuracy at the milliarcsecond level. By comparison, our results show significantly better agreement with predicted values based on historical VLBI observations than that with previous timing observations, as pulsar astrometry with the VLBI provides a more direct and model-independent method for accurately obtaining related parameters.

\par}\vskip 3mm
\normalsize\noindent{\narrower{PACS: 97.60.Gb; 95.85.-e; 95.85.Bh; 97.82.Cp}}\\
\noindent{\narrower{DOI: \href{http://dx.doi.org/10.1088/0256-307X/\cplvol/\cplno/\cplpagenumber}{10.1088/0256-307X/\cplvol/\cplno/\cplpagenumber}}

\par}\vskip 5mm


\section{Introduction}
Very Long Baseline Interferometry (VLBI) is a radio-astronomical technique that achieves high spatial resolution by utilizing antennas separated by thousands of kilometers (called a VLBI network). By now, the VLBI has consistently been the highest-resolution observational tool in astronomy, as it can provide an angular resolution much higher than the milli-arcsecond (mas) level. Due to its own radiation characteristics, pulsar research with the VLBI faces several challenges. Firstly, the pulsar is a kind of weak radio source with a flux density around the level of milliJanksy (mJy) at 1400 MHz. Secondly, the pulsar is a steep-spectrum source that is best observed at relatively low frequencies, making it more affected by the ionospheric effects. Thirdly,  the signal from a pulsar is a series of narrow radio pulses, which are quite distinct from the radiations of conventional radio sources. The history of pulsar studies with the VLBI can be traced back to some pathfinder observations with short-baseline radio interferometry in the 1970s \ucite{alp75,bs76}. Limited by some technical reasons (no effective methods to lengthen the coherent integration time, reduce the ionospheric effects, capture pulse signals, etc.), progress in pulsar VLBI observations proceeded slowly over the following two decades, with only some tentative observations \ucite{brs+85}.

Benefiting from the steady progresses of the VLBI observation, correlation, and data processing techniques since the 1990s, there have been tens of pulsars with their astrometric parameters obtained with the VLBI \ucite{cbs96, fgb99, bbg02, dtb07, cbv09, ysy13}. Though the distance and proper motion are fundamental and important pulsar parameters, it is challenging to obtain these astrometric parameters precisely. Except for a few pulsars (or their companion stars) with optical radiation, timing and VLBI are the two primary methods to measure astrometric parameters of pulsars. As most pulsars show timing irregularities (timing noise or glitches), it is also difficult to accurately fit their astrometric parameters with pulsar timing. By comparison, pulsar astrometry with the VLBI is more efficient and direct way, as this method can accurately track the motion trajectory of target pulsar on the sky by high-resolution imaging and only needs to fit five parameters, including the right ascension (RA) and declination (DEC) components of the proper motion ([$\mu_{\rm \alpha}$, $\mu_{\rm \delta}$]), annual parallax ($\Pi$) and position at a reference time ([$\rm RA_{\rm 0}$, $\rm DEC_{\rm 0}$]). Based on accurate astrometry results of pulsars, a series of studies have been conducted on important fundamental astrophysical questions, such as supernova explosion kick, equation of state of neutron stars, and Galactic electron density distribution. Accurate astrometry results obtained with the VLBI can improve the pulsar timing analyses \ucite{lys20}, so they are also useful to the pulsar timing arrays used for detecting nanohertz Gravitational waves \ucite{tyz13, xcg23}. Benefiting from the high accuracy and independence of pulsar astrometric parameters obtained with the VLBI, PSR J0337$+$1715 was also successfully identified as a millisecond pulsar stellar triple system by pulsar timing \ucite{rsa+14}. Considering the present situation that distance of more than 90\% pulsars are roughly estimated based on their dispersion measures (DMs) and Galactic electron density distribution models (such as TC93 \ucite{tac93}, NE2001 \ucite{cor02} and YMW16 \ucite{ymw17}), and DM$\textrm{-}$based distances of some individual pulsars may have great error \ucite{dtbr09} or systematic bias \ucite{lfl+06}, it is urgently needed to carry out observations with the VLBI on more pulsars. Furthermore, the inhomogeneity of interstellar medium (ISM) on a spatial scale as small as 0.05~AU was investigated by  100~$\mu$as resolution VLBI imaging of anisotropic interstellar scattering towards the target pulsar \ucite{bmg10}.

It is worthy of note that the majority of pulsar research works referenced above were achieved by the Very Long Baseline Array (VLBA), the European VLBI Network (EVN), and the Long Baseline Array (LBA). The Chinese VLBI Network (CVN), which consists of 5 antennas (Sheshan 25~m, Nanshan 26~m, Kunming 40~m, Miyun 50~m and Tianma 65~m that is also abbreviated as TMRT) and one data processing center in Shanghai Astronomical Observatory, has played an important role in some of the deep-space exploration projects (such as Chang'E series lunar probe missions and Tianwen Mars mission) \ucite{zzw19}. In comparison, there have been relatively few pulsar observations with the CVN, except for several tentative observations \ucite{gzz10, cjl16}. Some domestic researchers have applied for international VLBI observation time and have also obtained some interesting results \ucite{ysy13, dyc14}. With the ongoing construction of more radio telescopes in China, especially the world's largest and most sensitive radio telescope -  Five-hundred-meter Aperture Spherical radio Telescope (FAST) \ucite{nlj11,jiang20}, it's time to conduct more tentative pulsar observations with the CVN to enhance the network's capability to obtain more intriguing research results in the future.

In the following parts of this paper, we will present pathfinding pulsar observation results obtained with the CVN incorporating the FAST. The details of our observations and data reductions are described in Section 2 and Section 3, respectively. In Section 4, we will present the results of our observations and give further discussions.

\section{OBSERVATION AND DATA CORRELATION}
Figure~1~(a) shows the main steps of our observations and follow-up data correlations in a diagram. Normal pulsars usually show more irregular timing properties and higher velocities than millisecond pulsars \ucite{lor08}, so VLBI observations of normal pulsars need more sophisticated pulsar gating to catch their pulse signals and larger searching areas to get their positions. Considering these reasons, we chose two normal pulsars, PSRs~B0919+06 (J0922+0638) and B1133+16 (J1136+1551), as the targets of our pathfinding observations in order to test relevant technical routes. We use the B1950 names of target pulsars, which are well-known to researchers, but all their astrometry parameters (including historical results) used in this paper are in the J2000 equatorial coordinate. The typical periods of these two pulsars are 0.43~s and 1.19~s, respectively. Considering their power-law spectra and ionospheric effects, our observations were carried out in the L-band. Since the Kunming~40 m facility lacks an L-band receiver, only four telescopes listed in Table~1 joint our observations. Their common frequency range (1352$-$1480~MHz) was used and divided into four channels with a bandwidth of 32~MHz in our observations.

The observations were carried out on MJD~60045 (April 11th, 2023) with the phase-referencing mode. Two compact extra-galactic radio sources, 0918+082 and 1139+160 (with flux densities about 100~mJy), were selected as the phase-reference calibrators for PSR~B0919+06 and PSR~B1133+16, respectively. The angular separations from both calibrators to target pulsars are about 1.5~$^\circ$. More detailed parameters of target pulsars and (or) phase-reference calibrators are presented in Table~2. During observations, the antennas swept back and forth between the target pulsar and the calibrator source. In each phase-referencing observation cycle, the scan time on the calibrator and pulsar was about 2~min and 3~min, respectively. For FAST, the on-source scan time was comparatively shorter because of the longer switching time. The total observation length for each pulsar and corresponding calibrator was about 2~hr.

\vskip 4mm
\fl{1}
\centerline{\includegraphics[width=0.445\textwidth, angle=0]{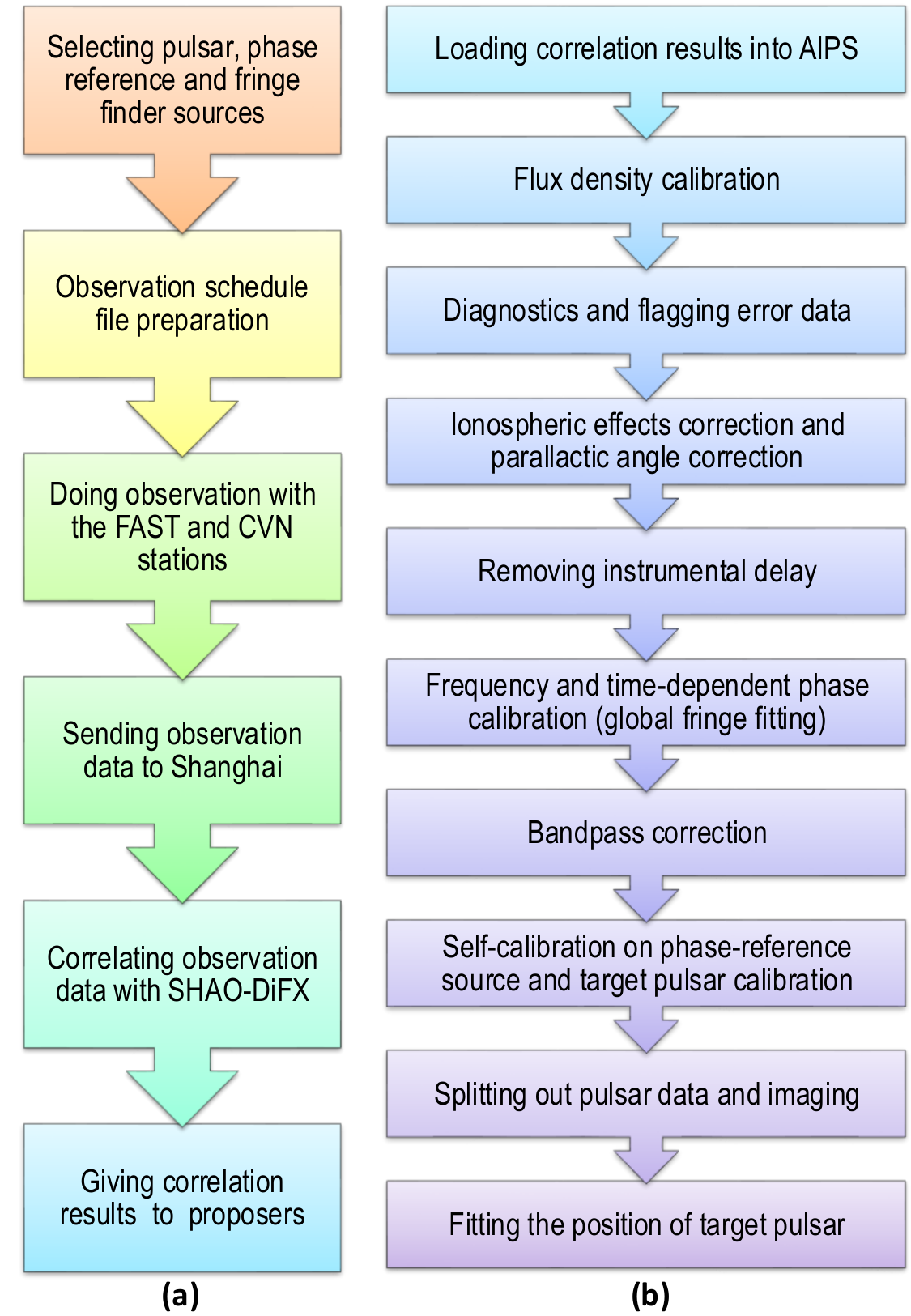}}
\vskip 2mm
\figcaption{7.5}{1}{(a) The flow diagram showing the main steps of our observations and follow-up data correlation; (b) the flowchart displaying the steps used in our data reduction procedure.}
\label{fig:diagram}
\medskip

\vskip 2mm
\tl{1}
\tabtitle{7.8}{1}{Parameters of telescopes used in our observations. The name, system equivalent flux density (SEFD), diameter, frequency coverage, and polarization of each telescope are presented from left to right. Dual-linear and dual-circular polarization is abbreviated as 2-LP and 2-CP, respectively.}\label{tab:cvn}
\vskip 2mm \tabcolsep 4.5pt
\centerline{\footnotesize
  \begin{tabular}{c c c c c c}
\hline\hline\hline
    Telescope & SEFD & Diameter & Frequency & Polarization \\
    (Name) & (Jy) & (m) & (GHz) & (CP/LP) \\
    \hline
    FAST \ucite{jiang20}& 0.9 & 500~m & 1.05-1.45 & 2-LP \\
    TMRT \ucite{yan24}  & 39 & 65~m & 1.35-1.75 & 2-CP \\
    Haoping \ucite{luo20} & 435 & 40~m & 1.1-1.75 & 2-CP \\
    Nanshan   & 300  & 26~m  & 1.0-2.0 & 2-CP\\
\hline\hline\hline
\end{tabular}}
\vskip 2mm

\vskip 2mm

\tl{2}\tabtitle{16.9}{2}{Parameters for target pulsars and phase-reference calibrators, including right ascension (RA), declination (Dec), RA error ($\rm E_{\rm RA}$), DEC error ($\rm E_{\rm DEC}$), mean flux density at 1400~MHz ($\rm S_{\rm 1400}$), typical flux density at C-band ($\rm S_{\rm C}$) and angular separation ($\theta_{\rm sep}$).}
\label{tab:psrcal}
\vskip 2mm \tabcolsep 4.5pt
\setlength{\tabcolsep}{3.5pt}
\centerline{\footnotesize
\begin{tabular}{ccccccccccccc}
\hline\hline\hline
\multicolumn{4}{c}{Pulsar}& & \multicolumn{6}{c}{Phase-referencing source}& &$\theta_{\rm sep}$ \\
\cline{1-4} \cline{6-11} \cline{13-13}
Name      & RA         & DEC          & $\rm S_{\rm 1400}$    &     & Name     & RA        & $\rm E_{\rm RA}$    & DEC     & $\rm E_{\rm DEC}$ & $\rm S_{\rm C}$ && Angle\\
(B1950)   & (hms)      & (dms)        & (mJy)    &     & (B1950)  & (hms)          & (mas)       & (dms)        & (mas)     & (mJy)       && ($^\circ$)\\
\hline
B0919+06 & 09:22:14.022 & +06:38:23.30  & $10\pm3$  & & 0918+082 & 09:21:01.0643 & 0.32 & +08:05:05.654 & 0.58  & $\sim74$  & & 1.48\\
B1133+16 & 11:36:03.1198 & +15:51:14.183 & $20\pm10$ & & 1139+160 & 11:42:07.7359 &  0.12 & +15:47:54.179 & 0.22 & $\sim126$ & & 1.46 \\
\hline
\hline\hline\hline
\end{tabular}}
\vskip 2mm

The collection of all sampled baselines on the (U,V) plane is called the UV-coverage. Figure~2 illustrates the UV-coverage during our observations on PSR~B0919+06 (top panel) and PSR~B1133+16 (bottom panel). It is evident that the length of the longest projected baseline in the East-West (U) direction is comparatively larger than that in the North-South (V) direction. From this, we infer that the observations will thus have higher resolution in right ascension (RA) than in declination (DEC).

\vskip 4mm
\fl{2}
\centerline{\includegraphics[width=0.445\textwidth, angle=0]{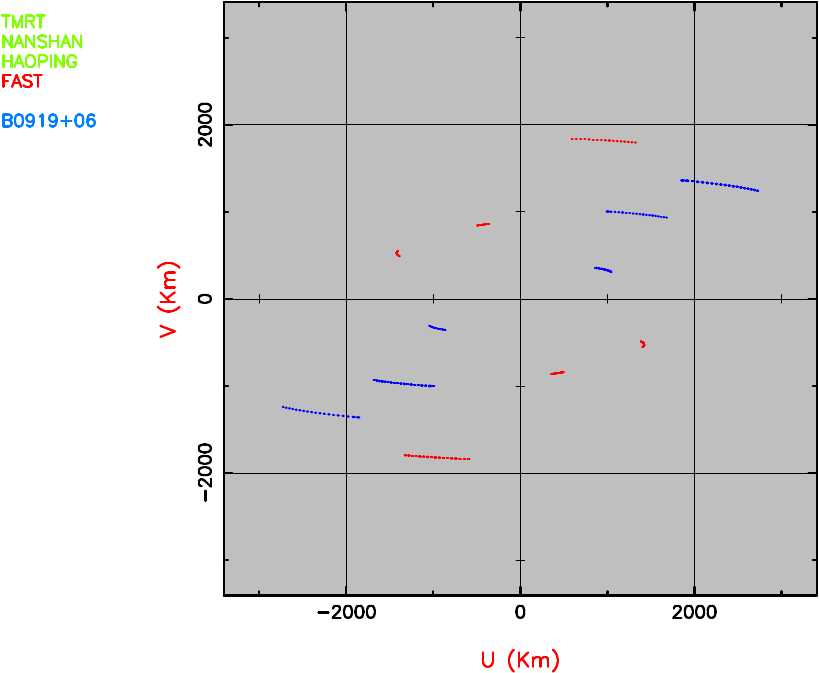}}
\centerline{\includegraphics[width=0.445\textwidth, angle=0]{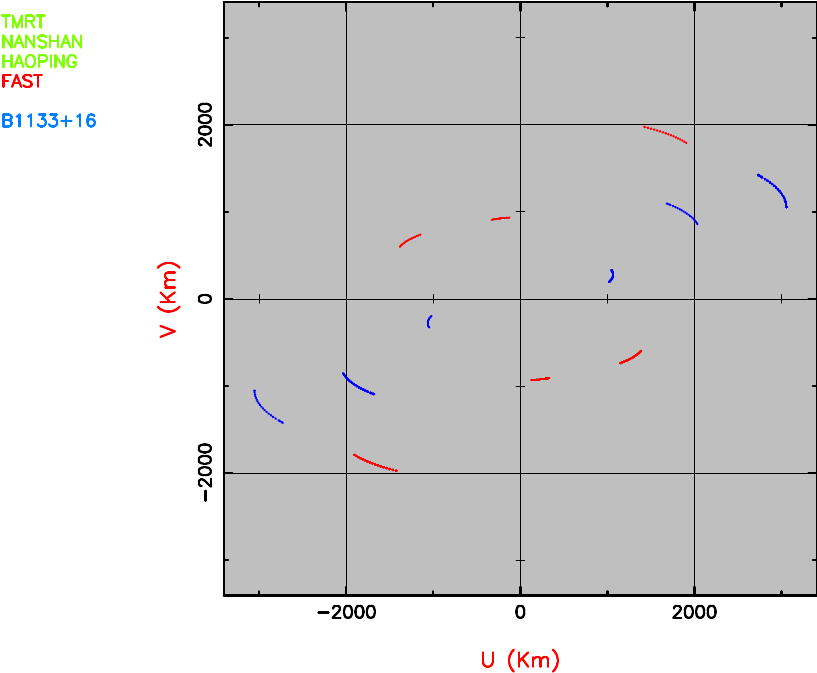}}
\vskip 2mm
\figcaption{7.5}{2}{The UV-coverage plots for our observations on PSR~B0919+06 (top panel) and PSR~B1133+16 (bottom panel). And the UV-coverage of baselines related to the FAST are shown with red curves. }
\label{fig:uvplt}
\medskip

In our observations, dual polarization data were recorded at all stations except Haoping. The observation data from each stations was transferred to Shanghai and correlated with the SHAO-DiFX correlator, which was built based on the free open-source distributed FX-style software correlator (DiFX) \ucite{dtb07, dbp11}. To effectively capture the pulse signal, pulsar binning was done on the observation data of target pulsars. The pulsar rotational phase was predicted using the pulsar timing software Tempo \ucite{nds15}. The ephemeris for target pulsars were obtained based on pulsar timing data from the TMRT. The correlated results were finally written out as FITS-IDI (FITS Interferometry Data Interchange) files.

\section{DATA REDUCTION}
As illustrated in Figure~1~(b), the data reduction was conducted using the NRAO Astronomical Image Processing System (AIPS) \ucite{gre03}, in accordance with the standard data processing procedures for L-band phase-referencing observations. After the data file was loaded into the AIPS, we initially conducted a preliminary check and flagged the data that was affected by serious radio frequency interferences (RFIs) or showed obvious errors. The JPL ionospheric model provided by NASA Jet Propulsion Laboratory was used to correct for phase variations caused by the ionosphere \ucite{wac99, sha99}. The parallactic angle and instrumental phase effects were subsequently corrected. As a phase-referencing observation, fringe fitting solutions on the phase-reference calibrator were applied to itself and the corresponding target pulsar. The bandpass effects were corrected with a strong radio source - OJ287 (unresolved flux density about 1.3~Jy). After that, data on the phase-reference source and target pulsar were averaged and split out for further analysis. Self-calibration solutions on a phase-reference source were further used on the target pulsar. Finally, the clean map of the target pulsar was generated with the robustness parameter around 0, balancing both resolution and image sensitivity considerations \ucite{bss99}. The average SEFD value for each telescope (obtained previously with other methods, see Table~1) was used to do the amplitude estimation, as there were no Tcal (system temperature calibration) signals injected during our observations at some stations. According to the theory of imaging with the VLBI, the visibility amplitude provides the brightness information of target source, while the phase tells its structure and position \ucite{mb08}. As we are interested in the positions of target pulsars, which are ideal point sources, our measurements are almost unaffected by the amplitude estimation method. The AIPS task JMFIT, which fits a two-dimensional Gaussian profile to a point source, was used to obtain the positions of target pulsars.

\section{Results and Discussions}
The target pulsars B0919+06 and B1133+16 were successfully detected by our observations. In Figure~3, the image plots for PSR~B0919+06 and PSR~B1133+16 are displayed on the top and bottom panels, respectively. The main beam of each observation is presented in the bottom left corner of the corresponding panel. As the pulsar is a very compact radio source, each target pulsar appears as a bright point in the central part of the sub-figure with the same shape as the synthesized beam. In the equatorial coordinate system, the current positions (on MJD~60045) of these two pulsars are [$\rm RA=09:22:14.0306485\pm0.000096262$, $\rm DEC=06:38:24.894337\pm0.00185165$] (for B0919+06) and [$\rm RA=11:36:03.0637488\pm0.000058898$, $\rm DEC=15:51:18.264981\pm0.00160380$] (for B1133+16). The position fitting errors in the DEC direction are larger than in the RA direction due to the relatively short baseline length (see Figure~2).

\vskip 4mm
\fl{3}
\centerline{\includegraphics[width=0.44\textwidth, angle=0]{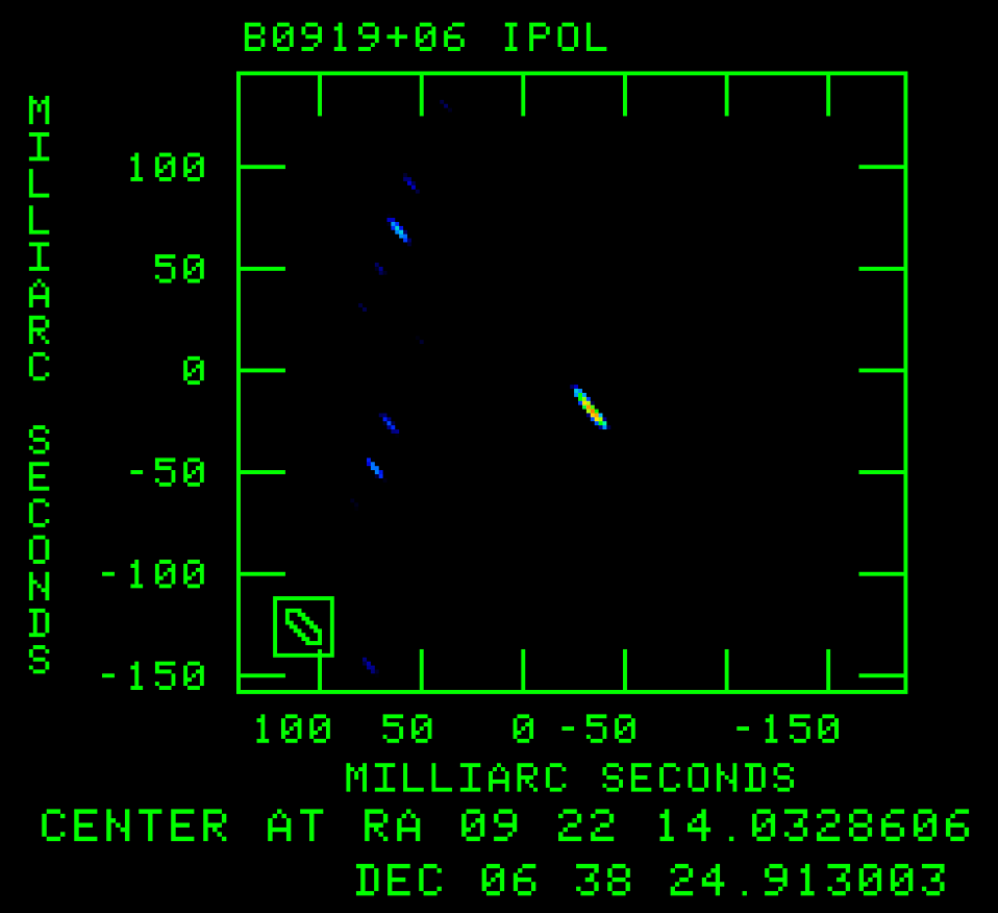}}
\centerline{\includegraphics[width=0.44\textwidth, angle=0]{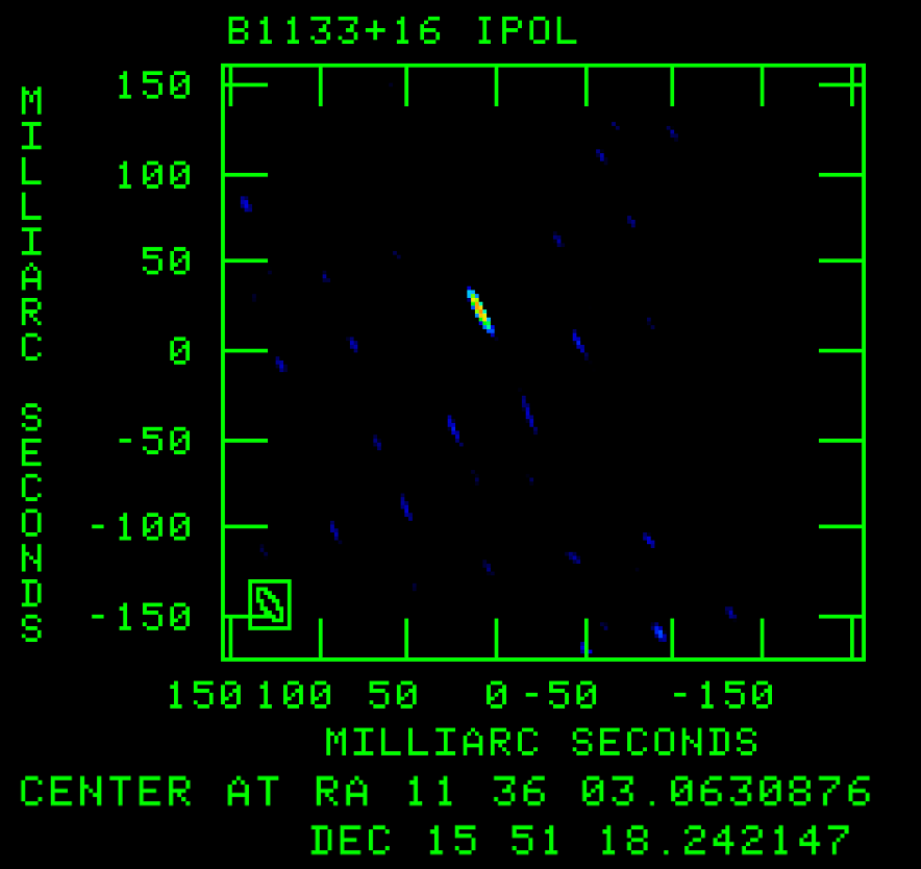}}
\vskip 2mm
\figcaption{7.5}{3}{The image of PSR~B0919+06 (top panel) and PSR~B1133+16 (bottom panel) obtained by us. The synthesized beam of observation on each pulsar is also plotted in the bottom left of corresponding sub-panel.}
\label{fig:image}
\medskip

\vskip 2mm
\tl{3}\tabtitle{17.0}{3}{Historical astrometry observation results and predicted position (on MJD~60045) of target pulsars.}
\label{tab:comphist}
\vskip 2mm \tabcolsep 4.5pt
\centerline{\footnotesize
\setlength{\tabcolsep}{1.5pt}
\begin{tabular}{ccccccccc}
\hline\hline\hline
Pulsar  & RA  & DEC  & Epoch  & $\mu_{\alpha}*cos\delta$  & $\mu_{\delta}$ & Method &$\rm RA^{i}$ & $\rm DEC^{i}$  \\
(B1950)     & (hms) & (dms) & (MJD) & (mas/yr)                & (mas/yr)     &  (tool) &  (hms) & (dms)   \\
\hline
\multirow{3}{*}{B0919+06} & 09:22:14.00423(1) & +06:38:22.8938(4) & 51525.1 & 18.35(6) & 86.56(12) & VLBA\ucite{ccl01} & 09:22:14.03291(9) & +06:38:24.9131(28) \\
         & 09:22:14.008(7) & +06:38:22.70(10) & 51361.9 & 18.8(9) & 86.4(7) & VLA\ucite{bfg03} & 09:22:14.0380(68) & +06:38:24.754(101) \\
         & 09:22:13.998(14) & +06:38:22.1(6) & 52264.0 & 21(14) & 28(34) & Nanshan\ucite{zhw05} & 09:22:14.028(24) & +06:38:22.70(94) \\
\hline
\multirow{2}{*}{B1133+16} & 11:36:03.1829(10) & +15:51:09.7257(150) & 51544.0 & -73.95(38) & 368.05(28) & VLBA\ucite{bbg02} & 11:36:03.06354(117)
 & +15:51:18.2926(164) \\
          & 11:36:03.1198(1) & +15:51:14.183(1) & 56000.0 & -73.785(21) & 366.569(64) & VLBA\ucite{dgb19} & 11:36:03.06309(10)
 & +15:51:18.2437(20) \\
\hline\hline\hline
\end{tabular}}
\vskip 2mm
\medskip

Table~3 lists the historical astrometric results of PSRs~B0919+06 and B1133+16, which were obtained by the interferometric method and the pulsar timing. The pulsar name, coordinate ([RA, DEC]), corresponding epoch (MJD), proper motion along RA $\&$ DEC ([$\rm {\mu_{\alpha}*cos\delta}$, $\rm {\mu_{\delta}}$]), measuring method, and predicted coordinate on MJD~60045 ([$\rm RA^{i}$, $\rm DEC^{i}$]) are listed sequentially from left to right in the columns of this table. For the interferometric method, the maximum baseline length of the VLA (Very Large Array) and VLBA is about 36~km and 8600~km, respectively. For the timing method, the proper motion of target pulsar was measured from regular timing observations using the Nanshan 25~m (upgraded to 26~m in 2015) telescope between 2000 January and 2004 August.

\medskip
\vskip 2mm
\fl{4}
\centerline{\includegraphics[width=0.45\textwidth, angle=0]{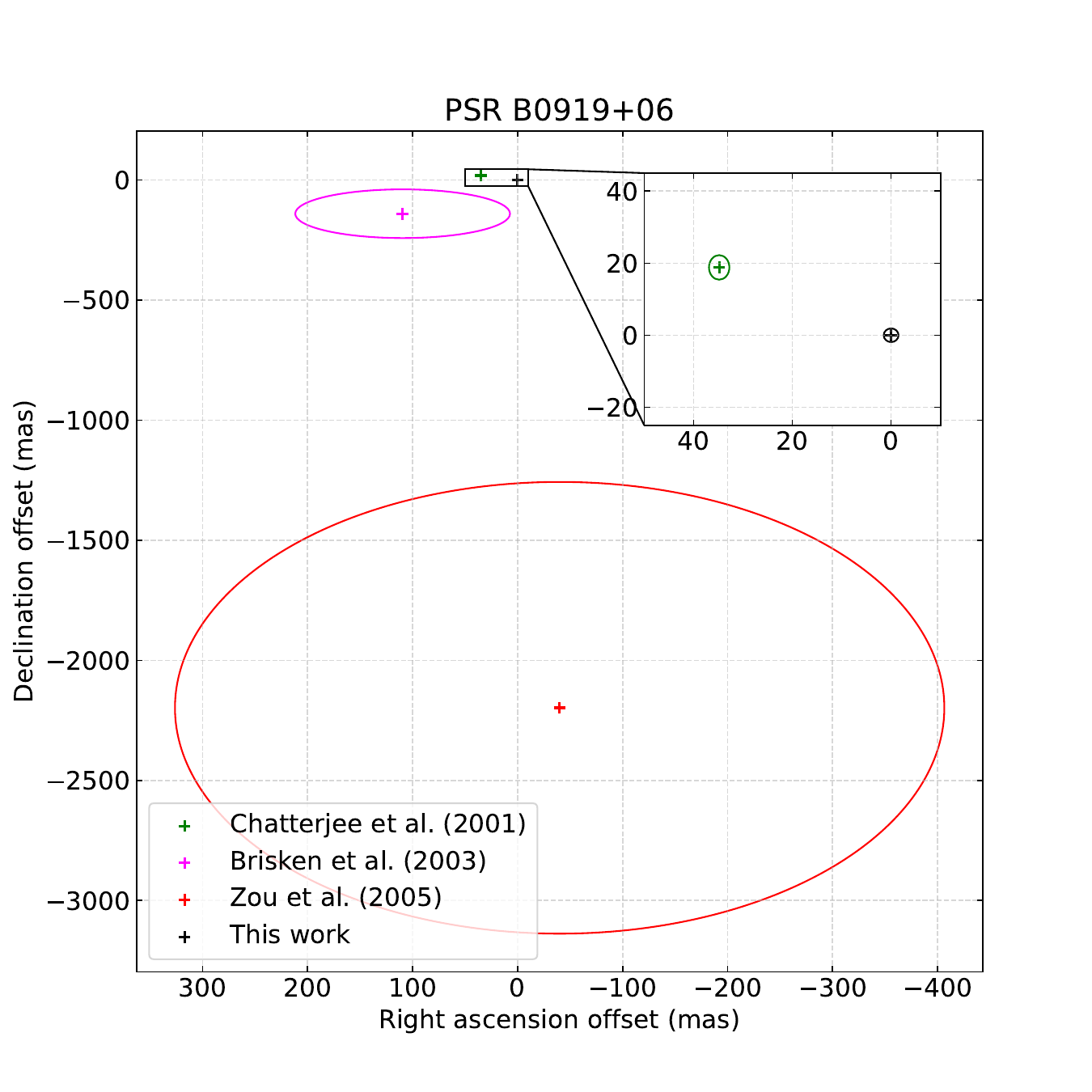}}
\centerline{\includegraphics[width=0.45\textwidth, angle=0]{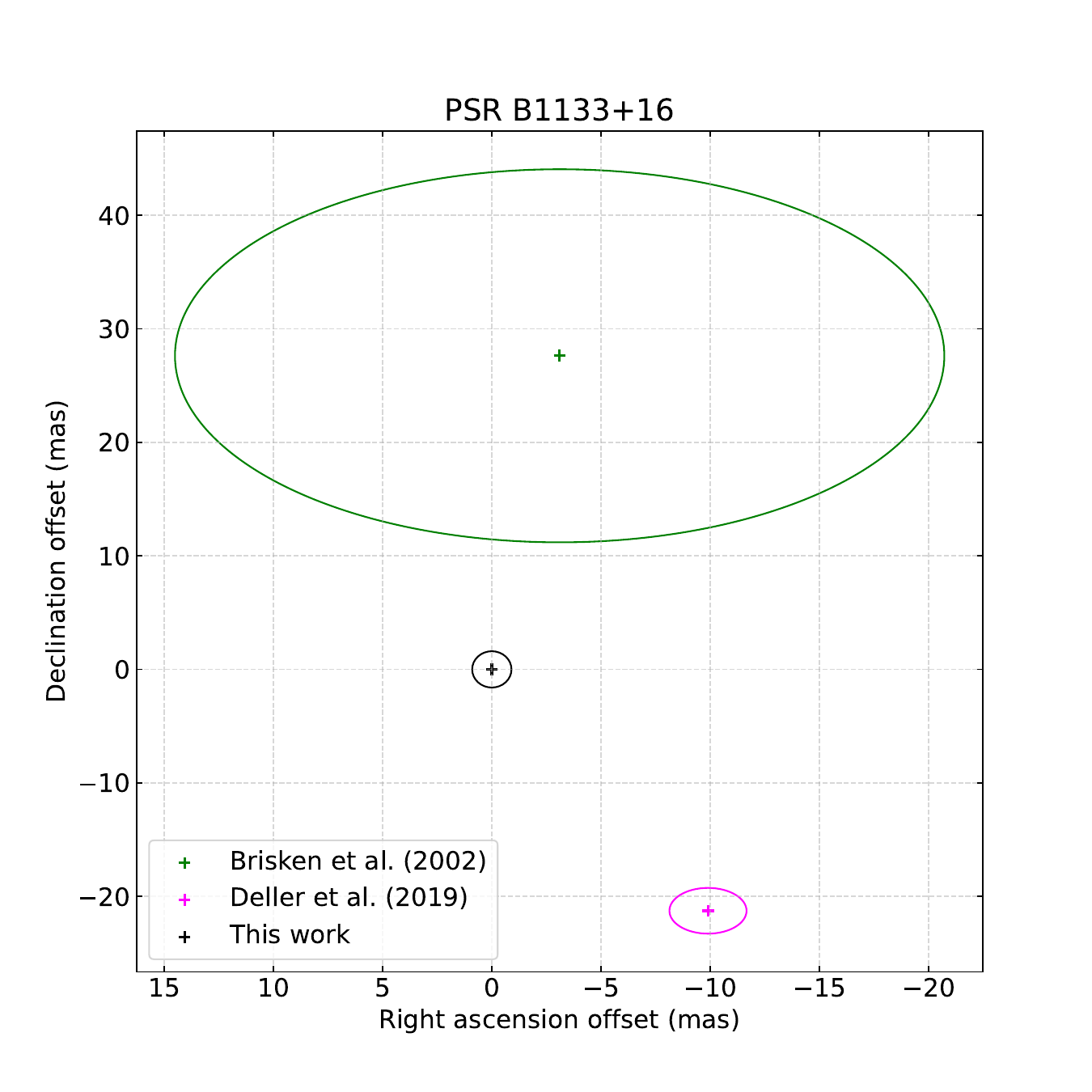}}
\vskip 2mm
\figcaption{7.5}{4}{Comparison plots about the predicted positions and our measurement results of PSR~B0919+06 (top panel) and PSR~B1133+16 (bottom panel). }
\label{fig:comphisnow}
\medskip

To make it easier to compare, we show contrasting plots of the predicted positions and our measurement results of PSR~B0919+06 and PSR~B1133+16 in the top and bottom panels of Figure~4. And for each pulsar, the current position that we have measured is set as the zero reference point. Taking into account the errors (1~$\sigma$) in the proper motions and annual parallaxes of target pulsars, the possible ranges are marked with ellipses around the best-predicted positions.

For PSR~B0919+06, it is clear that there are significant differences between the predicted positions obtained from the timing results and the interferometric results, while the differences among interferometric results are much smaller. Compared the position predicted from the VLA results by Brisken et~al.~(2003), our measurement results show slighter differences from the predicted position with VLBA results. The VLBA astrometry of PSR~B0919+06 was also performed in the L-band by Chatterjee et~al.~(2001), but they used different phase-reference sources and strategies. They chose 0906+015 (about $6^{\circ}$ away) and J0914+0245 (about $4^{\circ}$ away) as nodding calibrators in 1994$-$1996 and 1998$-$2000, respectively. In addition, they used J0922+0638 (a faint source about 12' away) as the in-beam calibrator source.

For PSR~B1133+16, its DEC measured by us is approximately in the middle of the predicted results based on the measurements by Brisken et~al.~(2002) and Deller et~al.~(2019) using the VLBA, but its RA position obtained by us is on the left of the predicted values with these two measurements. Brisken et~al.~(2002) and Deller et~al.~(2019) used J1143+1834 (about $3.25^{\circ}$ away) and 1139+160 (about $1.5^\circ$ away) as the phase-reference source for PSR~B1133+16, respectively. Compared to Brisken et~al.~(2002), the angular difference between the position of PSR~B1133+16 obtained by us and the prediction of Deller et~al.~(2019) is not obviously smaller, though we used the same phase-reference source as Deller et~al.~(2019).

Comparing our position measurement results of these two normal pulsars with previous works, it is clear that interferometric measurements can provide more accurate astrometric parameters of normal pulsars than timing observations. The main reason for this is that the data fittings of normal pulsars are affected by their irregular timing properties, such as timing noise and glitches. In contrast, pulsar astrometry with the interferometry is completely unaffected by irregular timing properties and can provide extremely precise results in a model-independent manner.

Compared with our measurement results with predicted values based on pulsar VLBI astrometry observations, we find that the differences are less than 50~mas, despite some of these historical pulsar astrometry results being obtained more than 20 years ago. There are many factors that can cause these differences, such as systematic offsets between different VLBI networks, residual ionospheric effects, different phase-reference calibrators, and structure evolutions of phase-reference sources \ucite{bcr96, rh14}. Judging from the systematic offset of about 1.5~mas found between the EVN and VLBA measurements \ucite{ysy13}, there may be a systematic offset of several mas level between the CVN and VLBA, as the positions of some new CVN  telescopes have not been accurately measured with the VLBI method. The astrometric accuracy achievable for an observation of an angular separation $\theta_{\rm sep}$  from a phase-reference source can be approximately estimated using the following equation:
\begin{equation}
\Delta_{\rm rel}\approx\theta_{\rm sep}\frac{c\Delta\tau}{|B_{\rm max}|}
\end{equation}
where $c$ is the speed of light,  $\Delta\tau$ is residual time delay, and $|B_{\rm max}|$ is the maximum baseline length.  $\Delta\tau$ normally comprises different types of error, such as residual troposphere delay $\Delta\tau_{\rm tropo}$, ionospheric delay $\Delta\tau_{\rm iono}$, structure evolution of phase-reference source $\Delta\tau_{\rm struc}$, etc. Considering our observations are carried out in the L-band, tropospheric effects are minimal, while ionospheric effects become more important. As what we mentioned previously, we tried to use the JPL ionospheric model to remove the ionospheric effects in our observation. Assuming that there are still 30\% of  ionospheric effects left,  the residual ionospheric delay ($c \Delta\tau_{\rm iono}$, in path length unit) can be calculated by the relation
\begin{equation}
|\it{c} \Delta\tau_{\rm iono}|={\rm 40.3} \big(\frac{\rm 0.30 \it I_{\rm e}}{\rm {TECU}}\big)\big(\frac{\nu}{\rm {GHz}})^{\rm {-2}}  (\rm {cm})
\end{equation}
where $\nu$ is the observation frequency, $I_{\rm e}$ is the total electron content (TEC), and 1 $\rm TECU$ corresponds to an electron column density of $10^{16}$~$\rm m^{-2}$ \ucite{rh14}.  During our observation, the TEC on the sky of FAST station changed most obviously from about 104 to 59~TECU with the observation time, while it changed least obviously from about 51 to 26~TECU at the Nanshan station. Meanwhile, $B_{\rm max}$ is about 3245~km and  $\theta_{\rm sep}$ is about $1.5^{\circ}$ in our observations on these two target pulsars. In our estimation about $|\it{c} \Delta\tau_{\rm iono}|$, if we use $I_{\rm e}=50$~TECU which is a medium  value for all stations in our observation, the corresponding position error caused by residual ionospheric effects should be about 5~mas. If we take the higher weight of FAST in the imaging process into consideration, residual ionospheric effects could contribute to the position error at the level of about 10~mas. In accordance with the VLBA Scientific Memo No.~18, when comparing the position of PSR~B0919+06 phase referencing to a nodding calibrator $4^{\circ}$ away to that obtained using an in-beam calibrator, it was found that their difference which was mainly caused by residual ionospheric effects could be as large as about 20~mas \ucite{cha99}. As for the effects of different phase-reference calibrators and their structure evolutions, we simply compared their positions used in previous research works with their current coordinates given in the Radio Fundamental Catalog (RFC, updated in 2024)\footnote{\url{http://www.astrogeo.org/calib/search.html}}, and found some obvious differences. There is a position difference [$\Delta_{\rm RA}=-0.66$, $\Delta_{\rm DEC}=1.70$]~mas for J0914+0245 compared to its position used in Chatterjee et~al.~(2001) with its present coordinate, while the  [$\Delta_{\rm RA}$, $\Delta_{\rm DEC}$] of 0906+015 is [5.88, 0.43]~mas. For J1143+1834 used in Brisken et~al.~(2002), we can also find a position difference of [$\Delta_{\rm RA}=-14.10$, $\Delta_{\rm DEC}=14.77$]~mas from its current coordinate.  And, there is also a position difference of [$\Delta_{\rm RA}=1.17$, $\Delta_{\rm DEC}=-2.05$]~mas for 1139+160 used in Deller et~al.~(2019) compared with that used in this work. Although the astrometric accuracies of these phase-reference sources were not as accurate as present measurements, their structure evolutions are not negligible for their positional differences from current results. Beside the referencing positions, the historical proper motion results of target pulsars were also affected by these factors. Even if the potential error in proper motion was as small as 0.5~mas/yr, the uncertainty in corresponding predicted position 20 years later would be about 10~mas. At present, we still cannot exactly figure out how these factors contribute to the differences between our measurement results and predicted values. But, we think that our measurement results align with historical research, as those differences are explainable if we take these errors into consideration.

Although we have successfully detected these two target pulsars and fitted their positions, which are consistent with historical VLBI measurements within reasonable ranges, there is a series of works to do if we want to obtain more important achievements with the CVN including FAST. As shown in Figure~2, the side lobe effect caused by sparse UV-coverage should be overcome by incorporating more telescopes or extending observation time. More standardized Tcal measurement solutions for VLBI observations should be accomplished at some stations. More special time should be arranged to look for in-beam calibrators for target pulsars like the PSR$\Pi$ project, which aims to finish the parallax ($\Pi$) and proper motion measurements of about 60 pulsars with the VLBA \ucite{dtbr09}. In addition, if a more accurate ionosphere model can be derived, it will be very helpful for phase-referencing precision in pulsar observations with the VLBI. Besides the present CVN telescopes, several large antennas are either planned or under construction, like the Qitai radio telescope (110~m) \ucite{wxm23}, the Jingdong pulsar radio telescope (120~m) \ucite{wxw22}, the radio telescope at Xiushui (120~m), the radio telescope at Rikaze (40~m), and the radio telescope at Changbaishan (40~m). As previously predicted \ucite{zb17, czj20}, more and more interesting research achievements about pulsars will be made with the rapidly growing CVN.

\medskip
\textit{Acknowledgements.} We would like to take this opportunity to acknowledge anonymous reviewers for their good suggestions. This work was supported by the National SKA Program of China (No. 2020SKA0120104, 2020SKA0120200), the National Key R\&D Program of China (No. 2022YFA1603104), and the National Natural Science Foundation of China (No. 12041301). Students Han Zhang and Zongming Zhang gave kind help on checking the English expressions in the article.

\end{document}